\documentclass[aps,pre,preprint]{revtex4}
\usepackage{graphicx}
\usepackage{amsmath}
\usepackage{amssymb}

\def\R{ {\mathbb R} }

\def\re{{\rm e}}
\def\ri{{\rm i}}
\def\rd{{\rm d}}
\def\fr {\mbox{$\frac{1}{2}$}}
\def\frr{\mbox{$\frac{1}{4}$}}

\begin{document}

\title{Enhancement of the Benjamin-Feir instability with dissipation}
\author{T.J. Bridges$^1$ and F. Dias$^2$}
\affiliation{$^1${\it Department of Mathematics,
University of Surrey, Guildford, Surrey GU2 7XH, England}\\
$^2${\it CMLA, ENS Cachan, CNRS, PRES UniverSud,
61, avenue du President Wilson, 94230 Cachan cedex, France}
}
 
\begin{abstract}
It is shown that there is an overlooked mechanism whereby
some kinds of dissipation can enhance the Benjamin-Feir instability
of water waves.  
This observation is new, and although it is counterintuitive, 
it is due to the fact that the Benjamin-Feir instability involves the
collision of modes with opposite energy sign (relative to the carrier wave),
and it is the \emph{negative energy perturbations} which are enhanced.\\

\end{abstract}

\maketitle

The discovery of the Benjamin-Feir (BF) instability of
travelling waves was a milestone in the history of water waves.  
Before 1960 the idea that a Stokes wave could be unstable does
not appear to be given much thought. The possibility that the Stokes wave could be unstable was
pointed out in the early 1960s \cite{phillips60,hasselmann62,lighthill65,whitham65}, 
but it was the seminal work of Benjamin and Feir 
\cite{bf67,benjamin67} which combined experimental evidence with a weakly
nonlinear theory that convinced the scientific community. 

Indeed, Benjamin \& Feir started their experiments in 1963 
\emph{assuming that the Stokes wave was stable}.  After several
frustrating years watching
their waves disintegrate -- in spite of equipment and laboratory
changes and improvements -- they finally came to the conclusion
that they were witnessing a new kind of instability.  The
appearance of ``sidebands'' in the experiments suggested the form
that the perturbations should take.  A history of these
experiments and the outcome is reported in \cite{tbb_memoir}. 

The theory of the BF instability is based on inviscid
fluid mechanics, and the assumption that the system is conservative.
Therefore it is natural to study the implication of perturbations on
the system.  The implications of a range of perturbations on the BF instability
have been studied in the literature: for example 
the effect of wind \cite{blh,bhl86} and the effect of viscosity
\cite{davey,blh,fabrikant,shhlps,yue}.
Some perturbations have been shown
to stabilize and others destabilize the BF instability.

\begin{figure}
\begin{center}
\includegraphics[angle=0,height=4.5cm]{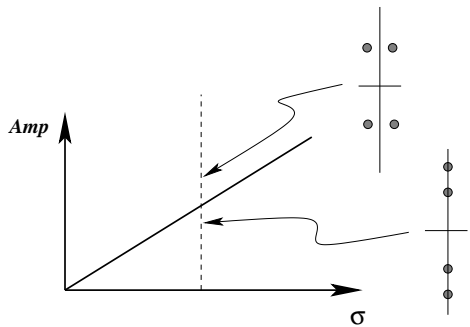}
\caption{\label{fig1} Schematic of the eigenvalue movement
associated with the BF instability, as a function
of the amplitude of the basic carrier wave ($Amp$) and the 
sideband wavenumber $\sigma$. For a fixed $\sigma$ (vertical dashed line), there is a threshold amplitude. 
Below the threshold, the eigenvalues are purely imaginary. At the threshold a collision occurs and above 
the threshold, the eigenvalues are complex.}
\end{center}
\end{figure}

However, there is a fundamental overlooked mechanism in all this work.
Mathematically, the BF instability can be characterized as a collision
of two pairs of purely imaginary eigenvalues of opposite energy sign as shown in Figure \ref{fig1}.   
In \cite{bm95}, this observation is implicit but the demonstration and
implications have not been given heretofore. This characterization of the BF instability also
appears in the nonlinear Schr\"odinger (NLS) model for modulation of
dispersive travelling waves \cite{ostro,zakharov}.
The eigenvalue with smaller positive imaginary part in the figure -- just before
collision -- has negative energy, whereas the
eigenvalue with larger imaginary part has positive energy.  This energy
is relative to the energy of the carrier wave $E^{{\rm Stokes}}$:
$E_-<E^{{\rm Stokes}}<E_+$ where $E_{\pm}$ are the energies of the modes
associated with the respective purely imaginary eigenvalues
in Figure \ref{fig1}.  Hence ``negative energy'' means that
$E_- - E^{{\rm Stokes}}<0$.

Once these facts are established we can appeal to the
result that dissipation can \emph{destabilize}
negative energy modes \cite{cairns,craik,mackay}.  
There are many examples in fluid mechanics where negative energy
modes -- which are stable in the inviscid limit -- are destabilized
by the addition of dissipation \cite{ort}: Kelvin-Helmholtz
instability \cite{craik,cairns,fs98}, interaction of a fluid with a
flexible boundary \cite{benjamin63,landahl}, stability of a fluid-loaded
elastic plate \cite{peake}, 
Euler modes perturbed by the Navier-Stokes equations \cite{dr98}.

The book of Fabrikant \& Stepanyants \cite{fs98} reports on experimental results for interfacial waves near
the Kelvin-Helmholtz instability illustrating the enhancement of the instability of
negative energy waves due to dissipation. See Figure 3.5 on page 83,
and the discussion on pp. 82-83 in \cite{fs98}.

In this brief communication, we sketch the basic result for water waves and then use
a nonlinear Schr\"odinger model perturbed by dissipation for illustration. The NLS equation has
shortcomings (e.g. symmetry which enables the phase to be factored out, lack of validity for
all time \cite{Craig}) but it provides a simple example of the phenomenon.

Davey \cite{davey} gives a general 
argument for the form of a dissipation-perturbed NLS model, and Blennerhassett
\cite{blh} starts with the full Navier-Stokes equations for a
free-surface flow with viscous free-surface boundary conditions
and derives a similar perturbed NLS equation.
For the dissipatively perturbed Stokes wave in deep water,
these NLS models take the form 
\begin{equation}\label{daveys_eqn1}
\ri A_t + (\alpha-\ri a) A_{xx} + \ri b A + (\gamma+\ri c)|A|^2A = 0\,,
\end{equation}
where $A$ is the envelope of the wave carrier,
and the modulations are restricted to one space dimension $x$.
When $a=b=c=0$, equation (\ref{daveys_eqn1}) reduces to the NLS equation
for the modulations of Stokes
waves in deep water; hence $\alpha$ and $\gamma$ are positive real numbers.
This NLS model has a BF instability and one can show explicitly that it involves a
collision of eigenvalues of the form shown in Figure \ref{fig1}.
We show below that when $a>0$, there is always
dissipation induced instability (before the BF instability), no matter
how small $a$ is.   The parameter $a$ is the perturbation of the
rate of change of the group velocity $\rd c_g/ \rd k$ due to dissipation.

First, consider the linear stability problem for gravity waves in
deep water. As the wave amplitude increases we show that there is a threshold value
at which two eigenvalues of the linear stability problem collide, and these two modes have
negative and positive energy.

With $\theta = x-ct$, the speed $c$ and amplitude $\eta$ of the
basic gravity wave of wavelength $2\pi/k$, to leading order, are
$c=c_0(1+k^2\varepsilon^2+\cdots)$, $c_0^2=gk^{-1}$,
\[
\eta(\theta)
= \varepsilon\eta_1(\theta) + \varepsilon^2\eta_2(\theta)
+\mathcal{O}(\varepsilon^3)\,,
\]
where $\varepsilon$ is a measure
of the amplitude,
\[
\eta_1(\theta) = \sqrt{2}\cos(k\theta-\theta_0)\,,\quad
\eta_2(\theta) = k\,\cos(2k\theta-2\theta_0)\,,
\]
with $\theta_0$ an arbitrary phase shift.
Using standard results on integral properties of Stokes waves,
the total energy \emph{relative to the moving frame} is
\[
E^{{\rm Stokes}} = T+V-cI = V-T\,,\quad\mbox{using}\quad 2T=cI\,,
\]
where $T$ and $V$ are the kinetic and potential energies respectively,
and $I$ is the momentum \cite{lh75}.  Substitution of the Stokes
expansion shows that $E^{{\rm Stokes}} =
0 + \varepsilon^3 E_3 + \mathcal{O}(\varepsilon^4)$.
Although the actual value of $E_3$ is not important for the argument
below, it is noteworthy that it is negative, and, using Table 2
of \cite{lh75}, one can confirm that $E^{{\rm Stokes}}$
is negative at finite amplitude as well.

To formulate the linear stability for gravity waves take
\[
\eta(\theta,x,t) \mapsto  \widehat\eta(\theta,\varepsilon)
+ \eta(\theta,x,t)\,,
\]
where $\widehat\eta(\theta,\varepsilon)$ is the basic carrier wave.
Take a similar expression for the velocity potential $\phi(\theta,x,y,t)$, where $y$
denotes the vertical space dimension.
Next one substitutes this form into the water wave equations, linearizes
about the carrier wave and takes $\eta(\theta,x,t)$ of the form
\[
\eta(\theta,x,t) = {\rm Re}\left( \Sigma(\theta,\sigma)
{\rm e}^{\ri \sigma x + \lambda t}\right)\,,
\]
where $\sigma$ is real (the modulation wavenumber), and
$\Sigma(\theta,\sigma)$ is periodic of the same period as the Stokes wave.
The result is an eigenvalue problem for the eigenfunction $\Sigma$ and
eigenvalue $\lambda$. 

The BF instability corresponds to a solution of this eigenvalue problem
with $0<\sigma \ll 1$ and ${\rm Re}(\lambda)>0$.  When $\sigma$ is fixed
-- but nonzero and small -- and the amplitude of the Stokes wave is
increased, there is a threshhold amplitude where the BF instability
occurs, and it corresponds to a collision of two eigenvalues
as shown in Figure \ref{fig1}.
To leading order the eigenvalues collide at
$\lambda = \pm \ri c_g\sigma$, where $c_g=\fr\sqrt{g/k}$ is the group velocity.

To show that the colliding modes have opposite energy sign, we need
a definition of the energy of the perturbation.  This definition
requires some consideration because the perturbation is
quasiperiodic in space: $2\pi/k-$periodic in $\theta$ and
$2\pi/\sigma-$periodic in $x$.  The total energy relative
to the moving frame is
\[
E^{{\rm total}} = \frac{\sigma}{2\pi}\int_0^{2\pi/\sigma} \frac{k}{2\pi}
\int_0^{2\pi/k}\, \widehat E\, \rd\theta\,\rd x\,,
\]
where $\widehat E=\widehat T-\widehat V - c\widehat I$,
$\widehat V = \fr g \eta^2$,
\[
\widehat T = \int_{-\infty}^{\eta}\fr(\phi_\theta^2+2\phi_\theta\phi_x +
\phi_x^2 + \phi_y^2)\rd y \,,\quad\mbox{and}\quad
\widehat I = \int_{-\infty}^{\eta} (\phi_\theta + \phi_x)\rd y\,.
\]
Evaluating the perturbation energy for the two modes that collide leads to
$E^{{\rm total}} = E^{{\rm Stokes}} + \varepsilon^2 E_2^{\pm} +\cdots$, with
\[
E_2^{\pm} = 2(k\pm\sigma)\left( 1 - \sqrt{1\mp \frac{\sigma}{k}}\right)
|{\rm C}_\pm|^2\,.
\]
Here ${\rm C}_{\pm}$ are scale factors associated with the eigenfunctions.
Clearly ${\rm sign}(E_2^+E_2^-) < 0$ for $0<\sigma \ll 1$.

Having shown that the colliding modes have opposite energy sign, we consider
a simple example which illustrates the mechanism for 
destabilization of negative energy modes by damping. A
prototype for a conservative system,
where the linearization has a collision of eigenvalues of opposite
energy sign, which is perturbed by Rayleigh damping is
\begin{equation}\label{model_dissip}
{\bf q}_{tt} + 2b{\bf J}{\bf q}_t + (\chi-\tau^2){\bf q} 
+ 2\delta{\bf q}_t={\bf 0}\,,\quad {\bf q}\in\R^2\,,
\quad {\bf J}= \begin{pmatrix} 0 & -1 \\ 1 & \hfill 0 \end{pmatrix}\,,
\end{equation}
where $\tau>0$ is the ``gyroscopic coefficient'', 
$\chi$ a real parameter with $|\chi| \ll \tau^2$, and $\delta\geq 0$.  

The energy of the system (\ref{model_dissip}) 
is strictly decreasing when $\delta>0$ and $\|{\bf q}_t\|>0$. 
Let ${\bf q}(t) = \widehat {\bf q}\re^{\lambda t}$;
then substitution into (\ref{model_dissip}) leads to the roots
\begin{equation}\label{C_roots}
\lambda = \ri \tau - \delta \pm\ri\sqrt{\chi+2\ri \tau\delta -\delta^2}\quad
\mbox{and}\quad
\lambda = -\ri \tau - \delta \pm\ri\sqrt{\chi-2\ri \tau\delta -\delta^2}\,.
\end{equation}
When $\delta=0$ there are four roots $\lambda = \pm\ri(\tau\pm \sqrt{\chi})$.
The eigenvalue movement shown in Figure \ref{fig1} is realized as $\chi$ decreases
from a positive value to a negative value, the collision occurring at $\chi=0$.
Suppose now $\chi$ is small and positive (just before the collision) and look
at the effect of dissipation on the two modes $\lambda_0 = \ri \tau \pm\ri \sqrt{\chi}$.
Substitution of the eigenfunctions for these two eigenvalues
into the energy shows that the mode associated with
$\ri \tau-\ri \sqrt{\chi}$ has negative energy while the mode associated with
$\ri \tau+\ri \sqrt{\chi}$ has positive energy.  

\begin{figure}
\begin{center}
\includegraphics[angle=0,height=3.0cm]{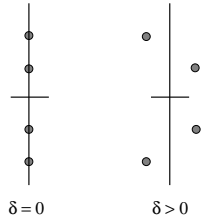}
\caption{\label{fig2} Schematic of the effect of dissipation on
the eigenvalues associated with (\ref{model_dissip}).}
\end{center}
\end{figure}

With $\delta$ small, expand the first pair of roots in (\ref{C_roots})
in a Taylor series
\[
\lambda(\delta) = \ri \tau\pm\ri \sqrt{\chi} \mp\frac{\delta}{\sqrt{\chi}}(\tau\pm \sqrt{\chi}) 
+ \mathcal{O}(\delta^2)\,.
\]
With $0<\delta\ll 1$ the eigenvalues are perturbed as shown
to the right in Figure \ref{fig2}. The negative energy mode,
$\lambda_0=\ri(\tau-\sqrt{\chi})$, has
positive real part when dissipatively perturbed, and the
positive energy mode, $\lambda_0=\ri(\tau+\sqrt{\chi})$, has negative
real part under perturbation.  Consequently, when small
dissipation is added to the otherwise \emph{stable} system (that is, $0<\chi \ll \tau^2$),
the mode with negative energy will destabilize. 
After the collision (when $\chi<0$) the growth rate of the instability is enhanced.  

It should be noted that other mathematically consistent forms
of damping can be used. For example the uniform damping
\begin{equation}\label{ud-eqn}
{\bf q}_t = \frac{\partial H}{\partial {\bf p}}- \delta{\bf q}\,,\quad
{\bf p}_t = -\frac{\partial H}{\partial {\bf q}} - \delta{\bf p}\,,
\end{equation}
makes mathematical sense. But it leads to uniform contraction of the phase space,
and does not destabilize negative energy modes. 

In order to study the effect of dissipation on water waves, one
could start with the Navier-Stokes equations
and perturb about the Stokes wave solution (see \cite{blh} for
instance for the case of
wind forcing).  Another approach is to add viscous perturbations to the potential
flow in various forms \cite{dd}. From the modified equations one can derive a
dissipative NLS equation.  There are two issues to highlight:
negative energy modes can be destabilized and so the BF instability can
be enhanced by dissipation, and secondly, the form of the damping is
important. It is known that negative energy modes of the Euler equations can be
destabilized by the form of damping found in the Navier--Stokes equations \cite{dr98}.

Following \cite{davey,blh}, a general perturbed NLS equation
for various types of physical situations can be written in the form (\ref{daveys_eqn1}).
The parameters $a$, $b$ and $c$ are taken to be non-negative. When they
are positive, they represent dissipative perturbations, since the
norm of the solution is strictly decreasing in
time when $a^2+b^2+c^2>0$.

When $a=b=c=0$, the resulting NLS equation is a Hamiltonian partial 
differential equation; with $A = u_1 + \ri u_2$ and ${\bf u}=(u_1,u_2)$,
\begin{equation}\label{nls-ham1}
{\bf J}{\bf u}_t = \nabla H({\bf u}) +
a{\bf J}{\bf u}_{xx} - b{\bf J}{\bf u} - c\|{\bf u}\|^2{\bf J}
{\bf u}\,,
\end{equation}
where ${\bf J}$ was defined in (\ref{model_dissip}) and
\begin{equation}\label{nls-energy}
H({\bf u}) = \int_{\R} \left[\,
\fr\alpha\|{\bf u}_x\|^2 -\frr\gamma\|{\bf u}\|^4\,\right]\,
\rd x\,.
\end{equation}

Let $\theta(x,t) = kx-\omega t+\theta_0$, and consider the basic
travelling wave solution to (\ref{nls-ham1}) when dissipation is
neglected,
\begin{equation}\label{TW}
\widehat {\bf u}(x,t) = {\bf R}_{\theta(x,t)}{\bf u}_0\,,\quad
{\bf R}_\theta =
\begin{pmatrix} \cos\theta & -\sin\theta \\
\sin\theta & \hfill\cos\theta \end{pmatrix}\,.
\end{equation}
Then ${\bf u}_0,\omega,k$ satisfy
$-\omega+\alpha k^2 = \gamma\|{\bf u}_0\|^2$.

It is assumed that the Stokes travelling wave exists for a 
sufficiently long
time before any dissipation can affect it: dissipation is taken to be
a second order effect.  

Next we check the energetics of the BF stability problem in NLS.
Linearize the partial differential equation (\ref{nls-ham1}) 
with $a=b=c=0$ about the basic travelling wave (\ref{TW}).
Letting ${\bf u}(x,t) = {\bf R}_{\theta(x,t)}( {\bf u}_0 + 
{\bf v}(x,t))$, substituting into the conservative version
of (\ref{nls-ham1}), linearizing about ${\bf u}_0$,
and simplifying leads to
\begin{equation}\label{lse_1}
{\bf J}{\bf v}_t 
+2\alpha k{\bf J}{\bf v}_x + \alpha {\bf v}_{xx}
 + 2\gamma\langle
{\bf u}_0,{\bf v}\rangle{\bf u}_0 = {\bf 0}\,,
\end{equation}
where $\langle\cdot,\cdot\rangle$ is the standard scalar
product on $\R^2$.

The class of solutions of interest are solutions which are periodic
in $x$ with wavenumber $\sigma$.  The parameter $\sigma$ represents
the sideband.  The BF instability will be associated with the limit
$|\sigma| \ll 1$.  Therefore let
\[
{\bf v}(x,t) = \fr {\bf v}_0(t) + \sum_{n=1}^{\infty}(
{\bf v}_n(t)\cos n\sigma x + {\bf w}_n(t)\sin n\sigma x )\,.
\]

Neglecting the $\sigma-$independent modes (superharmonic instability),
the $\sigma-$dependent modes decouple into $4-$dimensional
subspaces for each $n$, and satisfy
\begin{equation}\label{sigma_gen}
\begin{array}{rcl}
&&{\bf J}\dot{\bf v}_n + 2\alpha k n \sigma{\bf J}{\bf w}_n -
\alpha(n\sigma)^2{\bf v}_n + 2\gamma{\bf u}_0{\bf u}_0^T{\bf v}_n
= {\bf 0} \\
&&{\bf J}\dot{\bf w}_n - 2\alpha k n \sigma{\bf J}{\bf v}_n -
\alpha(n\sigma)^2{\bf w}_n + 2\gamma{\bf u}_0{\bf u}_0^T{\bf w}_n
= {\bf 0} \,.
\end{array}
\end{equation}
When the amplitude $\|{\bf u}_0\|=0$, it is easy to show that
all eigenvalues of the above system (i.e. taking solutions
of the form $\re^{\lambda t}$ and computing $\lambda$) are purely
imaginary.  Considering all other parameters fixed, and increasing
$\|{\bf u}_0\|$, we find that there is a critical amplitude
where the $n=1$ mode becomes unstable first through a collision of
eigenvalues of opposite signature. 

To analyze this instability, take $n=1$ and study the reduced
four dimensional system
\begin{equation}\label{four_dim}
\begin{array}{rcl}
&&{\bf J}\dot{\bf v}_1 + 2\alpha k \sigma{\bf J}{\bf w}_1 -
\alpha \sigma^2{\bf v}_1 + 2\gamma{\bf u}_0{\bf u}_0^T{\bf v}_1
= {\bf 0} \\
&&{\bf J}\dot{\bf w}_1 - 2\alpha k \sigma{\bf J}{\bf v}_1 -
\alpha\sigma^2{\bf w}_1 + 2\gamma{\bf u}_0{\bf u}_0^T{\bf w}_1
= {\bf 0} \,.
\end{array}
\end{equation}
To determine the spectrum, let $({\bf v}_1,{\bf w}_1) =
({\bf q},{\bf p})\re^{\lambda t}$. Then $(\lambda,\sigma)$
are determined by roots of
\[
\Delta(\lambda,\sigma) = \lambda^4 + 2(p^2+4k^2\alpha^2\sigma^2)\lambda^2
+(p^2-4k^2\alpha^2\sigma^2)^2 \,,
\]
where $p^2 = \alpha^2\sigma^4 - 2\alpha\gamma\|{\bf u}_0\|^2\sigma^2$.
Suppose $p^2>0$, then all four roots are purely imaginary (see Figure \ref{fig1}) and given by
\[
\lambda = \ri 2\alpha k\sigma \pm \ri p \quad\mbox{and}\quad
\lambda = -\ri 2\alpha k\sigma \pm \ri p \,.
\]
These modes are purely imaginary as long as $p^2>0$; equivalently
$2\gamma\alpha\|{\bf u}_0\|^2 < \alpha^2\sigma^2$. Since
$\alpha\gamma>0$, the instability threshold is
achieved when the amplitude reaches
\begin{equation}\label{thresh}
\|{\bf u}_0\| = \frac{|\alpha\sigma|}{\sqrt{2\alpha\gamma}}\,.
\end{equation}
At this threshold, a collision of eigenvalues occurs at the points
$\lambda = \pm 2\ri k \alpha\sigma$; see Figure \ref{fig1} for
a schematic of this collision.

It will be assumed henceforth that $k\neq0$.  Then instability
is through a collision of eigenvalues of opposite energy sign,
which reproduces the instability mechanism for the
full water-wave problem.

Purely imaginary eigenvalues of a Hamiltonian system have a
signature associated with them, and this signature is related
to the sign of the energy \cite{cairns,mack_saff,tjb97}.  Collision
of eigenvalues of opposite signature is a necessary condition
for the collision resulting in instability.

It is straightforward to compute the signature of the modes
in the NLS model. 
Suppose that the amplitude $\|{\bf u}_0\|$ of the basic state
is smaller than the critical value (\ref{thresh}) for instability.
Then there are two pairs of purely imaginary eigenvalues, and they
each have a signature.  Let us concentrate on the eigenvalues on the positive
imaginary axis
\begin{equation}\label{omega_def}
\lambda = \ri\Omega_\pm\quad\mbox{with}\quad
\Omega_\pm = c_g \sigma \pm p\,,\quad c_g = 2\alpha k\,.
\end{equation}
Then
\[
\textsf{Sign}(\Omega_\pm) = \ri\langle \overline{\bf q},{\bf J}{\bf q}\rangle
+\ri\langle \overline{\bf p},{\bf J}{\bf p}\rangle\,,
\]
where the inner product is real in order to make the conjugation
explicit.  One can also show that this signature has the same sign
as the energy perturbation restricted to this mode.
A straightforward calculation shows that
$\textsf{Sign}(\Omega_\pm) = \pm 4$ when $\|{\bf u}_0\|=0$.
Since $p^2$ decreases as the amplitude increases,
the two modes will have opposite signature for all $\|{\bf u}_0\|$
between $\|{\bf u}_0\|=0$ and the point of collision.

Now consider the effect of the damping terms.
Consider the reduced system (\ref{four_dim}) for the BF instability with
the $abc-$damping terms included:
\begin{equation}\label{four_dim_damping}
\begin{array}{rcl}
&&{\bf J}\dot{\bf v}_1 + 2\alpha k \sigma{\bf J}{\bf w}_1 -
\alpha \sigma^2{\bf v}_1 + 2\gamma{\bf u}_0{\bf u}_0^T{\bf v}_1
+ \mathcal{D}_1 = {\bf 0} \\
&&{\bf J}\dot{\bf w}_1 - 2\alpha k \sigma{\bf J}{\bf v}_1 -
\alpha\sigma^2{\bf w}_1 + 2\gamma{\bf u}_0{\bf u}_0^T{\bf w}_1
+ \mathcal{D}_2= {\bf 0} \,,
\end{array}
\end{equation}
with
\begin{equation}\label{damping_terms}
\begin{array}{rcl}
\mathcal{D}_1 &=& 2ka\sigma{\bf w}_1 + a\sigma^2{\bf J}{\bf v}_1
 + b{\bf J}{\bf v}_1+2c\langle{\bf u}_0,{\bf v}_1\rangle{\bf J}{\bf u}_0\\
\mathcal{D}_2 &=& -2ka\sigma{\bf v}_1 + a\sigma^2{\bf J}{\bf w}_1
 + b{\bf J}{\bf w}_1+2c\langle{\bf u}_0,{\bf w}_1\rangle{\bf J}{\bf u}_0\,.
\end{array}
\end{equation}
Now, let $({\bf v}_1(t),{\bf w}_1(t)) =
(\widetilde{\bf v}_1,\widetilde{\bf w}_1)\re^{\lambda t}$.  Then the
eigenvalue problem for the stability exponent reduces to studying the roots
of a determinant showing (with the help of \textsc{Maple}) that
the two roots in the upper half plane are given by
\begin{equation}\label{roots}
\lambda_{\pm} = 2\ri k\sigma\alpha - (b+a\sigma^2+c\|{\bf u}_0\|^2)
\pm\ri\sqrt{S}\,,
\end{equation}
with
\[
S = 4\ri a\sigma^3k\alpha-c^2\|{\bf u}_0\|^4-4k^2\sigma^2a^2-4\ri
a\sigma k\gamma\|{\bf u}_0\|^2 -2\alpha\gamma\sigma^2\|{\bf u}_0\|^2
+\alpha^2\sigma^4\,.
\]
When $a=b=c=0$, these stability exponents reduce to
\[
\lambda_{\pm} = 2\ri k\sigma\alpha \pm\ri\sqrt{\alpha^2\sigma^4 -2\alpha\gamma
\sigma^2\|{\bf u}_0\|^2}\,.
\]
Now suppose these two eigenvalues are purely imaginary: the
amplitude $\|{\bf u}_0\|$ is below the critical value
(\ref{thresh}).  To determine the leading order effect of
dissipation, expand (\ref{roots}) in a Taylor series with respect 
to $a,b$ and $c$ and take the real part
\begin{equation}\label{real_part}
{\rm Re}(\lambda_{\pm}) = -(a\sigma^2+b+c\|{\bf u}_0\|^2)\mp\frac{
2ak\sigma(\alpha\sigma^2-\gamma\|{\bf u}_0\|^2)}{
\sqrt{\alpha^2\sigma^4 -2\alpha\gamma
\sigma^2\|{\bf u}_0\|^2}} + \cdots\,.
\end{equation}
For any $a>0$ there is an open region of parameter space where these
two real parts have opposite sign since their product to leading order is
\[
{\rm Re}(\lambda_{-}){\rm Re}(\lambda_{+}) =
(a\sigma^2+b+c\|{\bf u}_0\|^2)^2-\frac{4a^2k^2\sigma^2(\alpha\sigma^2-\gamma
\|{\bf u}_0\|^2)^2}{\alpha^2\sigma^4 -2\alpha\gamma
\sigma^2\|{\bf u}_0\|^2}+\cdots\,.
\]
For any $a,b,c$ with $a\neq0$ there is an 
open set of values of $\|{\bf u}_0\|$ where this expression is strictly
negative, showing that ${\rm Re}(\lambda_-)$ and ${\rm Re}(\lambda_+)$
perturb in opposite directions.  In 
this parameter regime the dissipation perturbs the
negative energy mode as shown schematically in Figure \ref{fig2}.

It is clear that when only the $b-$term is present all eigenvalues shift to the left. Therefore the $b-$term does not
produce any enhancement of the instability, in agreement with \cite{shhlps}. This damping
is analogous to the \emph{uniform damping} in (\ref{ud-eqn}).
It is the $a-$term which 
leads to enhancement.  However the NLS is a simplified model for water waves.

In summary, the fundamental observation is that BF instability is associated with a
collision of eigenvalues of positive and negative energy, and there are physically 
realizable forms of damping which enhance this instability.  It remains to be seen how 
this effect can be revealed in laboratory experiments, in numerical experiments
based on the full water-wave equations, and in the open ocean.
 
\section*{Acknowledgements}
\noindent This work was enhanced by a grant of a CNRS Fellowship
to the first author, and by support from the CMLA at
Ecole Normale Sup\'erieure de Cachan.  Helpful discussions with Gianne
Derks are gratefully acknowledged.
%
%

\newpage
Figure 1. Schematic of the eigenvalue movement
associated with the BF instability, as a function
of the amplitude of the basic carrier wave ($Amp$) and the 
sideband wavenumber $\sigma$. For a fixed $\sigma$ (vertical dashed line), there is a threshold amplitude. 
Below the threshold, the eigenvalues are purely imaginary. At the threshold a collision occurs and above 
the threshold, the eigenvalues are complex.

Figure 2. Schematic of the effect of dissipation on
the eigenvalues associated with (\ref{model_dissip}).
\end{document}